\documentclass[a4paper,11pt]{article}
\usepackage[margin=1in]{geometry}
\usepackage{epsfig}
\usepackage{cite}
\usepackage{graphicx}
\usepackage{epstopdf}
\usepackage{color}
\UseRawInputEncoding
\usepackage[nottoc]{tocbibind}
\usepackage[english]{babel}
\usepackage{tabularx}
\usepackage{xparse}
\usepackage{braket}
\usepackage{amsmath}
\usepackage{hyperref}
\usepackage{grfext}
\usepackage{amssymb}
\usepackage{caption}
\usepackage{mathtools}
\usepackage{subcaption}
\usepackage{algorithm2e}
\usepackage{float}
\usepackage{subcaption}
\usepackage{multirow}
\restylefloat{table}
\usepackage{booktabs}
\usepackage[dvipsnames]{xcolor}
\usepackage{tikz}
\usepackage{pdflscape}
\usepackage{lscape}
\date{\today}
 \normalsize

\newcommand{\bmat}{\left(\begin{array}}
\newcommand{\emat}{\end{array}\right)}
\newcommand{\be}{\begin{equation}}
\newcommand{\ee}{\end{equation}}
\newcommand{\bea}{\begin{eqnarray}}
\newcommand{\eea}{\end{eqnarray}}

\def\gtwid{\mathrel{\raise.3ex\hbox{$>$\kern-.75em\lower1ex\hbox{$\sim$}}}}
\def\ltwid{\mathrel{\raise.3ex\hbox{$<$\kern-.75em\lower1ex\hbox{$\sim$}}}}
\def\gev{{\rm \, Ge\kern-0.125em V}}
\def\tev{{\rm \, Te\kern-0.125em V}}

\def    \be            {\begin{equation}}
\def    \ee            {\end{equation}}
\def    \bea           {\begin{eqnarray}}
\def    \eea           {\end{eqnarray}}
\def\eps{\epsilon}
\def\a{\alpha}
\def\b{\beta}
\def\g{\gamma}
\def\d{\delta}
\def\n{\nu}

\def\vphi{\varphi}

\def\m{\mu}

\def\d{\delta}

\def\m{\mu}

\def\d{\delta}

 \normalsize

\begin{document}

\vspace{.3cm}

\title{\Large \bf Palatini $f(R)$ gravity and variants of k-/constant roll/warm inflation within variation of strong coupling scenario}
\author
{  \it \bf M. AlHallak$^{1}$\thanks{mahmoud.halag@unitedschool.ae},  A. AlRakik$^{1}$\thanks{amer.alrakik@univ-tlse3.fr}, N. Chamoun$^{2}$\thanks{nidal.chamoun@hiast.edu.sy} and M. S. El-Daher$^{3,4}$\thanks{m-saemaldahr@aiu.edu.sy}
 \\\hspace{-3.cm}
\footnotesize$^1$  Physics Department, Faculty of Sciences, Damascus University, Damascus, Syria,  \\\hspace{-3.cm}
\footnotesize$^2$ Physics Department, HIAST, P.O. Box 31983, Damascus, Syria, \\\hspace{-3.cm}
\footnotesize$^3$  Higher Institute of Laser Research \& Applications, Damascus University, Damascus, Syria\\ \footnotesize$^4$ Arab International University, Daraa, Syria.
}

\date{}
\maketitle

\abstract{We show that upon applying Palatini $f(R)$, characterized by an $\a R^2$-term, within a scenario motivated by a temporal variation of strong coupling constant, then one gets a quadratic kinetic energy. We do not drop this term, but rather study two extreme cases: $\a <<1$ and $\a >>1$. In both cases one can generate a kinematically-induced inflationary paradigm. In order to fit the Planck 2018 data, the $\a >>1$ case, called k-inflation, requires a fine tuning adjustment with non-vanishing non-minimal coupling to gravity parameter $\xi$, whereas the $\a <<1$ case, studied in the  constant-roll regime, can fit the data for vanishing $\xi$. The varying strong coupling inflation scenario remains viable when implemented through a warm inflation scenario with or without $f(R)$ gravity.}



\section{Introduction}
In \cite{chamoun_ijmpa_2021}, we adopted a model of variations of constants in order to generate an inflationary scenario, where the strong coupling was assumed to vary in time encoded in a scalar field representing this variation. Although current geophysical and astronomical data preclude any variation of constants, be it strong coupling \cite{chamoun_plb_2001}, or Higgs vev \cite{chamoun_joP_2005}, or electric charge \cite{bekenstein}, however no data precludes variation in very early times. In \cite{chamoun_ijmpd_2008}, a connection between variation of constants and inflation was suggested, whereas in \cite{chamoun_jcap_2016} this idea was pursued further into a concrete model shown to be able to accommodate data in some variants involving multiple inflaton fields. Alternatively, the single inflaton model was shown in \cite{chamoun_ijmpa_2021} to be viable provided one changes the gravitational sector and assumes $f(R)$ gravity.

Usually, any model of inflation is defined by the choice of the scalar fields involved, their kinetic terms, mutual couplings and potentials, and couplings to gravity. However, we likewise have to
specify the gravitational action with the corresponding degrees of freedom.
One example of the latter is the choice between the metric and the Palatini formulations. The simplest extended gravitational action is given by replacing the Einstein-Hilbert action of general relativity (GR) by a function $f(R)$ of the Ricci scalar. Whereas both formalisms agree in GR, they do differ in $f(R)$ gravity.  $f(R)$ with metric formulation was studied extensively (see \cite{Metric} and references therein), whereas $f(R)$ in Palatini formalism constitutes a current hot topic, studied say in \cite{Palatini} and references therein.

In our inflationary model based on couplings time variation, the addition of an $\a R^2$ term in the pure gravity Lagrangian changed the potential into an effective one, but also led to a quadratic kinetic energy term which was dropped in \cite{chamoun_ijmpa_2021} on the ground that it involved an $\a$-coupling which could be argued to be small perturbatively, and this allowed to derive formulae for the spectral index $n_s$ and the scalar-to-tensor ratio $r$ contrasted with planck data 2018 \cite{Planck_2018} separately or combined with other experiments \cite{bicep2-keck_2015}. Actually, the model can be considered as a special case of \cite{Enckel} which treated the general case of an arbitrary potential leading also to a quadratic kinetic energy term. However, in our model the potential is not arbitrary but dictated from new physics linking the two concepts of ``inflation'' and `variation of constants". Thus, our setup models the variation of coupling by a scalar field with, according to Bekenstein arguments \cite{chamoun_plb_2001,bekenstein}, self coupling, and, furthermore, we assume an additional conformal invariant non-minimal coupling of the scalar field to gravity, which in turn is given by $f(R)$ (classically equivalent to tensor-scalar model) and not by GR.

The aim of this work is twofold. First, we study the effect of the quadratic kinetic energy term. For this, we take two extreme cases. The first case corresponds to $\a >> 1$, which makes the scalar field non-canonical per excellence. Many studies were carried out to refine the inflationary scenario within the framework of scalar fields possessing
a non-canonical kinetic term \cite{Unnikrishnan_jcap_2012,Odintsov, Karam}. Actually, such kinematically induced inflationary scenarios go back to Starobinsky model \cite{Starobinsky} more than four decades ago, which considered a geometrical modification to general relativity in order to explain inflation. Nonetheless, the Starobinsky model, when considered in the framework of the Palatini formalism, in contrast to the metric formulation, can not represent a model for inflation, due to the absence of a propagating scalar degree of freedom that can play the inflaton role \cite{Karam2}. Here, we go beyond and consider a scalar field, motivated by a non-geometrical origin suggested by variation of constants $\grave{a}$ la Bekenstein, minimally or nonminimally coupled to gravity with a potential whose form is dictated by Bekenstein arguments \cite{chamoun_plb_2001,bekenstein}. We find that with a non-vanishing non-minimal coupling to gravity (non-MCtG) the model can fit the data. However, one can not get closed forms of the ``canonical" potential except in some cases which we illustrate in order to show the ``plateau"-form of the potential in terms of the ``canonical" field which rolls slowly during inflation.

The second case corresponds to the perturbative regime where we restrict the analysis to first order in $\a$. Our model in this case parallels the well known constant-roll k-inflation \cite{Odintsov_2019}, and we prove that within a given limit corresponding to vanishing non-MCtG with $\a$ small, $\ell$ large the model is viable, and we check this numerically for both small and large constant-roll parameter $\beta$.

Inflationary scenarios by variation of constants generically suffer from appealing to new physics for an exit scenario during reheating \cite{chamoun_jcap_2016}. A solution to this problem is provided by warm inflation paradigm \cite{WarmInflation}. In this paradigm, the radiation era is accompanying the slow roll regime, and no need for an exit scenario. For this, our second objective is to add the warm inflation ingredient into our  varying coupling inflation scenario. We find that with no $f(R)$ gravity the solution is hardly viable, but with Palatini $f(R)$, which would correspond to new degrees of freedom, accommodation of data is easily met.

The paper is organized as follows. In section 2, we introduce the model and illustrate how the quadratic kinetic energy appears. In section 3, we study the case of large $\a$ computing the spectral parameters to be contrasted with data. Section 4 is devoted to the study of the ``canonical" potential shape when $\a>>1$.  In section 5, we analyse the perturbative regime where $\a$ is small, whereas in section 6 we prove its viability. In section 7, we treat the case of warm inflation in a certain weak limit. We end up with a summary and conclusion in section 8.

%%%%%%%%%%%%%%%%%%%%%%%%%%%%%%%%%%%%%%%%%%
\section{Analysis of the basic model}
\label{1}

Our starting point is the general four dimensional action:
\bea S = S_{\phi}+S_{g} + S_{g\phi}\eea
where $S_{\phi}$ is the varying strong coupling constant action given by \cite{chamoun_jcap_2016}:
\bea \label{action} S_{\phi}\equiv \int d^4x \sqrt{-g} \mathcal{L}_{\phi}  &=& \int d^4x \sqrt{-g} \lbrace -\frac{1}{2} f(\phi) g^{\mu \nu} \partial_\mu \phi \partial_\nu \phi -V(\phi) \rbrace\eea
 where $f(\phi)=\frac{1}{\ell^2 \phi^2}$, and $V(\phi)=\frac{V_0}{\phi^2}$ with $\phi$ embodying the strong coupling constant variation $g^{st}(x)=g^{st}_0  \phi(x)$ and $\ell$ is the Bekenstein length scale, and $V_0=\frac{\langle G^2\rangle_T}{4}$ encodes the gluon field strength vacuum expectation value (vev) at inflation temperature $T$,  whereas  $S_{g}$ is the pure gravity Lagrangian including the Einstein-Hilbert action to which is added an $f(R)$ gravity term taken in our case as a quadratic function of the Ricci scalar $\a R^2$, and we include also a coupling term $S_{g\phi}$ between gravity and the field $\phi$. Adopting units where the Planck mass $M_{pl}$ is equal to one, we have:
 \bea S_{g} &=& \int d^4x \sqrt{-g} \left[ \frac{1}{2} \left( R + \a R^2 \right)\right] \\ S_{g\phi} &=& \int d^4x \sqrt{-g} \left[ -\xi R \phi^2 \right] \eea
 with $R$  the Ricci Scalar constructed from the metric $g_{\mu\nu}$. Note that the form of the potential in Eq. (\ref{action}) is not put by hand, but rather is dictated by the physical assumption of a varying strong coupling constant, where gauge and Lorentz invariance impose this form originating from the gluon condensate \cite{chamoun_ijmpd_2008, chamoun_jcap_2016}.

 We start by making a change of variable absorbing the function $f$ in order to get a ``canonical'' kinetic energy term. Thus we introduce the field $h$ defined as $\phi=\exp(\ell h)$, so that to get the action:
\bea
\label{fr-action}
S= \int d^4x \sqrt{-g} \left[\frac{1}{2} F(R) + \frac{1}{2} G(h) R - \frac{1}{2} g^{\a\b} \partial_\a h\partial_\b h - V(h)\right]
\eea
where
\bea
V(h)=V_0 \exp(-2 \ell h), G(h)= -\xi \exp(2 \ell h), F(R) = R + \a R^2
\eea
Instead of using at this stage the $1^{st}$-order cosmological perturbation theory, by perturbing the metric ($g_{\m\n} \rightarrow g_{\m\n} + \d g_{\m\n}$) and keeping terms of first order in the perturbations, we anticipate that the $\a R^2$ would contribute involved terms upon this metric change, so we follow \cite{faraoni-0805.1726, Enckel} and introduce an auxiliary field $\psi$ and an action:
 \bea
\label{fr-action2}
S= \int d^4x \sqrt{-g} \left[ \frac{1}{2} G(h) R + \frac{1}{2} \left\{F(\psi) + F'(\psi) (R-\psi)\right\}  - \frac{1}{2} g^{\a\b} \partial_\a h\partial_\b h - V(h)\right]
\eea
The equation of motion of $\psi$ gives $R=\psi$ provided $F''(\psi) \neq 0$. We change variable again $\psi \rightarrow \lambda$ such that ($\lambda=F'(\psi)= 1+2\a \psi$), so we get
\bea
\label{fr-action3}
S= \int d^4x \sqrt{-g} \left[ \frac{1}{2} \left\{\lambda + G(h)\right\} R - \frac{1}{2} \left\{\psi \lambda -F(\psi(\lambda)) \right\}  - \frac{1}{2} g^{\a\b} \partial_\a h\partial_\b h - V(h)\right]
\eea
We carry out a conformal transformation on the metric
\bea
\label{fr-conformal}
g_{\a\b} \rightarrow \Upsilon^2 g_{\a\b}= \tilde{g}_{\a\b} &:& \Upsilon^2= \lambda+G(h)
\eea
then we get in the ``Metric" formulation, where the Christoffel symbols are defined in terms of the metric and thus are not independent and the corresponding affine connection is defined to be the Levi-Civita one, the following \cite{Budhi}:
\bea
\label{fr-action-metric}
S^{\mbox{\tiny ``Metric"}} &=& \int d^4x \sqrt{-\tilde{g}} \left[ \frac{1}{2} \tilde{R} - \frac{3}{4} \frac{\tilde{g}^{\m\n}}{(\lambda+G(h))^2} \nabla_\m (\lambda + G(h)) \nabla_\n (\lambda + G(h)) \nonumber \right. \\ && \left.    -  \frac{1}{2} \frac{1}{\lambda+G(h)} \tilde{g}^{\a\b} \partial_\a h\partial_\b h - \tilde{V}(h,\lambda)\right] \\
\tilde{V}(h,\lambda)&=& \frac{V(h)+W(\lambda)}{(\lambda + G(h))^2}
\label{tildeV}
\eea
where
\bea \label{W} W(\lambda)&=& \frac{1}{2} \left[ \psi \lambda - F(\psi(\lambda))\right]
= \frac{(\lambda - 1)^2}{8 \a}
\eea
 We see that in the ``Metric" formulation, we get a kinetic energy term for ($\lambda + G(h)$), and the field $\lambda$ is dynamic, i.e. its equation of motion cannot be solved algebraically.

 For simplicity, then,  we restrict the study from now on to the ``Palatini" formulation, where the Christoffel symbols are considered independent and are to be determined dynamically. Remembering here that the pure gravity is not represented by a simple $R$-term, then the connection will be different from the Levi-Civita one. Under this formulation, we get (noting that $ \sqrt{-{g}}=\Upsilon^{-4} \sqrt{-\tilde{g}}, g^{\a\b}= \Upsilon^{2} \tilde{g}^{\a\b}$ and $R = \Upsilon^{2} \tilde{R}$):
 \bea
\label{fr-action-palatini}
S^{\mbox{\tiny ``Palatini"}} &=& \int d^4x \sqrt{-\tilde{g}} \left[ \frac{1}{2} \tilde{R}    -  \frac{1}{2} \frac{1}{\lambda+G(h)} \tilde{g}^{\a\b} \partial_\a h\partial_\b h - \tilde{V}(h,\lambda)\right]
\eea
where again $\tilde{V}(h,\lambda)$  is given by Eq. (\ref{tildeV}), and where eq. (\ref{W}) is again valid.

The equation of motion of $\lambda$ can be solved algebraically to give it in terms of the field $h$ and its derivatives, so $\lambda$ is not a new degree of freedom
\bea
\label{varphi}
\lambda &=& \frac{1+G(h)+8\a V(h) +2 \a G(h) (\partial h)^2}{1+ G(h) -2\a (\partial h)^2}
\eea

Substituting (Eq. \ref{varphi}) in (Eq. \ref{fr-action-palatini}), we get (dropping the ``Palatini'' superscript and the $\sim$ over the metric):
\bea
\label{fr-action-palatini-almostfinal}
S &=& \int d^4x \sqrt{-g} \left[ \frac{R}{2} -  \frac{1}{2} \frac{1}{(1+G(h))(1+8\a \bar{U})} g^{\a\b} \partial_\a h \partial_\b h \right. \nonumber \\ && \left.
+\frac{\a}{2} \frac{1}{(1+G(h))^2(1+8 \a \bar{U})} (\partial^\a h \partial_\a h)^2
- \frac{\bar{U}}{1+8 \a \bar{U}}\right]
\eea
where
\bea
\label{barU}
\bar{U} &=& \frac{V(h)}{(1+G(h))^2} = \frac{V_0 \exp{(-2 \ell h)}}{(1-\xi \exp{(2 \ell h)})^2}.
\eea
In order to get a ``canonical'' kinetic energy term, we again make the change of variable ($h \rightarrow \chi$)  by
\bea
\label{h-chi}
\frac{dh}{d\chi} &=& \pm \sqrt{(1+G(h)) (1+8\a \bar{U})}
\eea
to get finally
\bea
\label{fr-action-palatini-final}
S &=& \int d^4x \sqrt{-g} \left[ \frac{R}{2} -  \frac{1}{2} g^{\a\b} \partial_\a \chi \partial_\b \chi
+\frac{\a}{2} (1+8 \a \bar{U}) (\partial^\a \chi \partial_\a \chi)^2
- U\right]
\eea
where
\bea
\label{Ueffective-h}
U &=& \frac{\bar{U}}{(1+8\a \bar{U})}=\frac{V_0}{8\a V_0 + \left(e^{\ell h} - \xi e^{3 \ell h}\right)^2}
\eea
We see here that the effect of $\a R^2$-term is manifested in two ways. First, it helps in getting a ``flat" effective potential $U$. Actually, regardless of the form of $\bar{U}$, we see that the $\a R^2$-term leads, say when $\bar{U} (V_0)$  increases in modulus indefinitely, to an effective potential with a flat portion ($U \sim (8\a)^{-1}$). Second, the $\a R^2$-term leads to the appearance of squared kinetic energy $(\partial^\a \chi \partial_\a \chi)^2$.

In \cite{chamoun_ijmpa_2021}, $\a$ was taken to be small in a way to neglect the quadratic kinetic energy term. In fact, upon perturbing the metric then the $(\a \d g)$-term would give higher order terms, whereas the $\a (\partial^\b \chi \partial_\b \chi)^2$ would give, in the slow-roll inflationary era, contributions of order $\a \dot{\chi}^4$ which is subdominant compared to the $\a$-correction in $U$. Thus, in \cite{chamoun_ijmpa_2021}, one could apply the shortcut ``potential method'', using $U$ as an effective potential. We intend now to refine this analysis, and consider the effect of the quadratic kinetic energy, keeping first order in $\a$ when $\a$ is small, and studying, in addition, the case where $\a$ is large. We point out here that although we do not present explicitly the Einstein/$f(R)$ field equations, however we use known formulae for the spectral observables ($n_S,r$) in the different limits under consideration, which were derived using first order cosmological perturbation theory in solving the field equations \cite{chamoun_ijmpa_2021}.

\section{k-inflation: case $\a>>1$ }
Under the assumption
\bea
\label{case1}
1&\ll& \a (1+8 \a \bar{U}) (\partial^\a \chi \partial_\a \chi)
\eea

our k-inflation model features a single scalar field with the action
\bea
\label{fr-action-palatini-final}
S &=& \int d^4x \sqrt{-g} \left[ \frac{R}{2}
+\frac{\a}{2} (1+8 \a \bar{U}) (\partial^\a \chi \partial_\a \chi)^2
- U\right]
\eea
Introducing the ``standard" field $\varphi$ defined by
\bea
\label{varphi-chi}
\frac{\partial \varphi}{\partial \chi} &=& \left[ 2 \a (1+8 \a \bar{U} ) \right]^{\frac{1}{4}}
\eea
we get a ``standard" form for the k-inflation Lagrangian
\bea
\label{fr-action-palatini-final}
\mathcal{L} =  \frac{R}{2} + p(\varphi,X)  &:& p(\varphi,X)= X^2 - U(\varphi),  X=\frac{1}{2} \partial^\a \varphi \partial_\a \varphi \\
S &=& \int d^4x \sqrt{-g} \left[ \frac{R}{2}
+(\frac{1}{2} \partial^\a \varphi \partial_\a \varphi)^2
- U\right]
\eea
The spectral index $n_S$ and the tensor-to-scalar ratio are given now by \cite{Liddle_jcap_2012}:
\bea
\label{k-inflation_parameters}
n_s-1 = \frac{1}{3} \left(4 \eta - 16 \eps \right) &,& r=16 c_s \eps
\eea
where
\bea
\label{eps-eta-cs}
\eps = \frac{1}{2} 3^{\frac{1}{5}} \frac{(U_{,\varphi})^\frac{4}{3}}{U^\frac{5}{3}} &,&
\eta = 3^\frac{1}{5} \frac{(U_{,\varphi})_{,\varphi}}{U^\frac{2}{3} (U_{,\varphi})^\frac{2}{3}} \\
&c_s^2 = \frac{p_{,X}}{2X(p_{,X})_{,X}+p_{,X}}= \frac{1}{3}&
\eea
where the comma ($_,$) means differentiation with respect to what follows it. However, one should note that in order to compute the derivative with respect to the ``standard'' field $\varphi$, one should differentiate $U$ with respect to $h$, which is known from Eq. (\ref{Ueffective-h}) and using Eqs. (\ref{h-chi},\ref{varphi-chi}), so to get
\bea
\label{derivatives}
\frac{dh}{d\varphi} &=& \left(\frac{(1-\xi e^{2 \ell h})^2+8\a V_0 e^{-2\ell h}}{2\a}\right)^\frac{1}{4} \\
U_{,\varphi} &=& \frac{dU}{dh}\frac{dh}{d\varphi},\\
(U_{,\varphi})_{,\varphi}&=& \frac{d^2U}{dh^2} (\frac{dh}{d\varphi})^2 + U_{,h} \frac{dh}{d\varphi} \frac{d}{dh}(\frac{dh}{d\varphi})
\eea
The input parameters are ($V_0, \ell, \a, \xi$) and the initial values of the ``original'' inflaton field $h$ at the start of inflation. However, one can show analytically that the model is able to fit the data for some regions in the parameter space, Actually, we get  the following analytic formulae
\bea
\label{scenarioII}
1-n_s =& \frac{4  {\ell}^{4/3} \left(3  {\xi} e^{4  {\ell}  {h}} (44  {\a}  {V_0} {\xi}-7)+e^{2  {\ell}  {h}} (2-112  {\a}  {V_0}  {\xi})+12  {\a}  {V_0}+21  {\xi}^4 e^{10 {\ell}  {h}}-59  {\xi}^3 e^{8  {\ell}  {h}}+57  {\xi}^2 e^{6  {\ell} {h}}\right)}{\sqrt[3]{ {\a}} \sqrt[3]{ {V_0}} \left(\left(3-3  {\xi} e^{2  {\ell}  {h}}\right) \left(3  {\xi} e^{2 {\ell}  {h}}-1\right)\right)^{2/3} \left(8  {\a}  {V_0}+ {\xi}^2 e^{6  {\ell}  {h}}-2 {\xi} e^{4  {\ell}  {h}}+e^{2  {\ell}  {h}}\right)^{2/3}}  \\
\mathrel{\stackrel{\makebox[0pt]{\mbox{\normalfont\tiny $y \equiv \xi e^{2\ell h}$}}}{=}}& \frac{  {\ell}^{4/3} \left(12 - 112 y + 132 y^2\right)}{\left(\left(3-3  {\xi} e^{2  {\ell}  {h}}\right) \left(3  {\xi} e^{2 {\ell}  {h}}-1\right)\right)^{2/3}} + \mathcal{O}\left(\frac{1}{\a V_0}\right)\\
\mathrel{\stackrel{\makebox[0pt]{\mbox{\normalfont\tiny $\xi=0$}}}{=}}& \frac{8  {\ell}^{4/3} \left(6  {\a}  {V_0}+e^{2  {\ell}  {h}}\right)}{3^{2/3} \sqrt[3]{ {\a}} \sqrt[3]{ {V_0}} \left(8  {\a}  {V_0}+e^{2  {\ell}  {h}}\right)^{2/3}}
\\
r= & \frac{16  {\ell}^{4/3} e^{2  {\ell}  {h}} \left(1- {\xi} e^{2  {\ell}  {h}}\right)^{4/3} \left(3 {\xi} e^{2  {\ell}  {h}}-1\right)^{4/3}}{\sqrt[6]{3} \sqrt[3]{ {\a}} \sqrt[3]{ {V_0}} \left(8 {\a}  {V_0}+ {\xi}^2 e^{6  {\ell}  {h}}-2  {\xi} e^{4  {\ell}  {h}}+e^{2  {\ell} {h}}\right)^{2/3}} \\
\mathrel{\stackrel{\makebox[0pt]{\mbox{\normalfont\tiny $\xi=0$}}}{=}}&
\frac{16  {\ell}^{4/3} e^{2 {\ell} {h}}}{\sqrt[6]{3} \sqrt[3]{{\a}} \sqrt[3]{{V_0}} \left(8 {\a} {V_0}+e^{2 {\ell} {h}}\right)^{2/3}}
\eea
Thus, enforcing the Bekenstein hypothesis $\ell >1$, which means in our adopted units the absence of any length scale shorter than Planck length, we see that for ($\xi=0$) one can not accommodate the data requiring ($0< 1-n_s << 1$ and $0<r<<1$). However, in the limit ($\a V_0 >>1$), one can adjust the parameter ($y = \xi e^{2 \ell h}$) around the roots of ($12-112+132y^2$) and get the data fit. Actually, the two roots ($0.125792,0.722693$) of the latter polynomial are less than one, which implies that $h$ at the start of inflation was negative. Physically, this means that the strong coupling $g^{st}$ was less than its current value ($\phi < 1$).

\begin{figure}[H]
\includegraphics[width=11.5 cm]{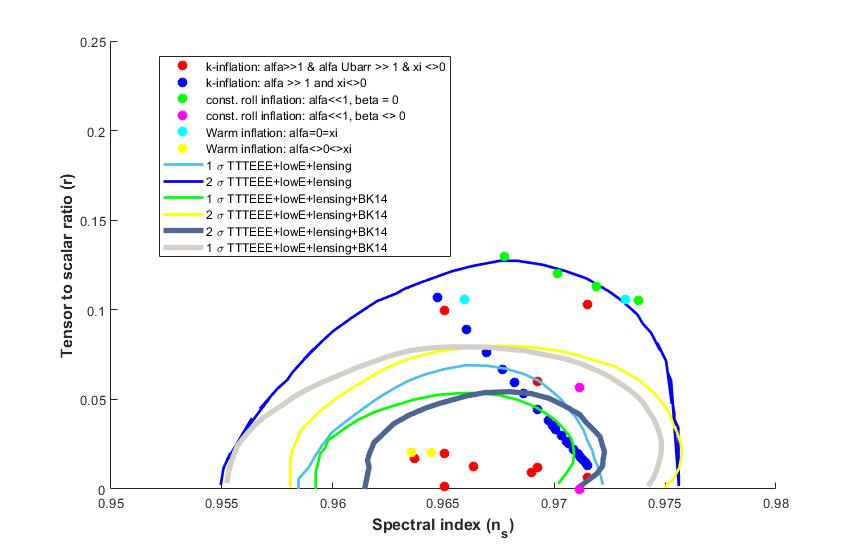}
\caption{ Kinematically derived inflation, in a model of varying strong coupling constant ($g^{st}(x)=g^{st}_0 \phi(x)$), with $f(R)$ gravity via $\a R^2$-term and non-MCtG $-\xi R \phi^2 $-term. The blue points correspond to the limit where $\a >>1$, and we imposed $\xi=1$. The red points correspond to the limit $\a>>1$ and $\a \bar{U}>>1$ with $\xi \neq 0$. The green (pink) points correspond to the limit $\a <<1$ with $\xi=0$ and a vanishing (non-vanishing) constant-roll parameter $\b$. The yellow (skyblue) points correspond to warm inflation scenario with (without) $f(R)$ gravity. The models are contrasted to Planck 2018, separately or combined with other experiments, contour levels of spectral parameters ($n_s,r$). All acceptable points correspond to $\ell > 1$. \label{fig1}}
\end{figure}

Fig. \ref{fig1}, indeed, shows that there are acceptable points, colored in blue, for the following scanning:
\bea
\label{scanning}
\ell \in [1,3], \a \in [10,20],  e^{2\ell h} \in [0.723031,0.723035], V_0 \in [0.5,2], \xi=1.
\eea

%%%%%%%%%%%%%%%%%%%%%%%%%%%%%%%%%%%%%%%%%
\section{The ``plateau'' shape: case $\a>>1$ }
From Eq. \ref{Ueffective-h} we see that in the limit where
\bea
\label{scenarioI}
1<<8\a \bar{U} &\Leftrightarrow& (1-\xi e^{2\ell h})^2 e^{2\ell h} << 8\a V_0
\eea
the effective potential shows a ``plateau'' form ($U(h)\sim \frac{1}{8\a}$), and our objective in this section is to study the shape of this plateau in terms of the ``canonical'' field $\varphi$ which is the field to roll slowly along the effective potential.

Actually, one would like, starting from the known potential $U(h)$ given in Eq. \ref{Ueffective-h}, to find an analytic expression of the potential in terms of the ``canonical'' field $\varphi$. However, it is not possible in general to do this, as we can not carry out analytically the following integral, originating from Eqs. (\ref{h-chi},\ref{varphi-chi}), let alone invert it to express $h$ in terms of $\varphi$:
\bea
\label{varphi-h}
\varphi &=& (2\a)^{1/4}\int \frac{dh}{\left[(1-\xi e^{2\ell h})^2+8\a V_0 e^{2\ell h}\right]^{1/4}}
\eea
Even in the case of MCtG ($\xi=0$), and although one can in principle carry out the above integration but the resulting expression involving hypergeometric functions is not invertible.

\begin{figure}[hbtp]
\centering
\begin{minipage}[l]{0.5\textwidth}
\includegraphics[width=7. cm]{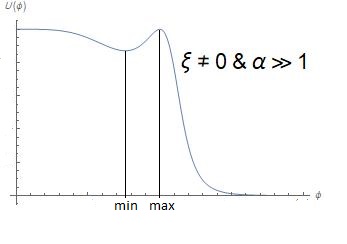}
\caption*{A. $max\equiv\varphi_0=(\frac{4}{V_0 \xi})^{1/4} \ell^{-1}$. $min\equiv \varphi_0 3^{-1/4}$}
\end{minipage}%
\begin{minipage}[l]{0.5\textwidth}
\includegraphics[width=5.5 cm]{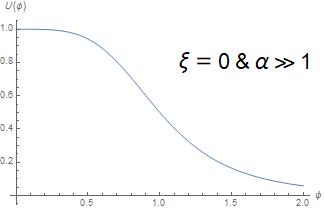}
\caption*{B.  $\xi=0$ }
\end{minipage}
\vspace{0.5cm}
\caption{Plateau shape in the limit of Eq.(\ref{scenarioI}). Scenario (B) with $\xi =0$ fits data provided $\ell <1$}
\label{fig2}
\end{figure}

However, in the limit of Eq. (\ref{scenarioI}), one can carry out analytically the integration and get
\bea
\label{scenarioI-varphi-h}
e^{\ell h} &=& \frac{\sqrt{V_0}}{2} (\ell \varphi)^2
\eea
and we see that the effective potential is given as
\bea
\label{effective-scenarioI}
U(\varphi) &=& \left[8\a+\frac{\ell^4 \varphi^4}{4} \left(1-\xi \frac{V_0}{4} \ell^4 \varphi^4\right)^2\right]^{-1}
\eea
In the left part (A) of Fig. \ref{fig2}, we plot the shape of effective potential, and find that it has one local maximum (minimum) at $\varphi_0=\sqrt[4]{4/(V_0 \xi)} \ell^{-1}$ ($\varphi_0/\sqrt[4]{3}$). We see that the limit of Eq. (\ref{scenarioI}) is equivalent to
\bea
\label{scanerioI-equivalentlimit}
\ell^2 \varphi^2 \left|1-(\varphi/\varphi_0)^4\right| &<<& \sqrt{32 \a}
\eea
Thus, we see that as long as the field, during its slow rolling along the plateau from $\varphi=0$, does not meet the local minimum, then the slow roll condition is satisfied and the inflationary solution is consistent. In the right part (B) of Fig. \ref{fig2}, we draw the same plateau in the case of MCtG. However, the solution is not viable for $\ell >1$.

As a matter of fact, one can compute the observable parameters ($n_s,r$) using the effective potential expression in this limit (Eq. \ref{effective-scenarioI}), and we find with the combination ($z=\xi V_0 \varphi^4 \ell^4$) the following
 \bea
\label{scanrioI-ns-r}
1-n_s&=& \ell^{4/3} 3^{-2/3}\frac{48-112z+33z^2}{(8-8z+3z^2/2)^{2/3}} \\
r&=& 2^{2/3} 3^{-1/6} 8\ell^{16/3} \varphi^4 \frac{(16-16z+3z^2)^{4/3}}{512\a + \ell^4 \varphi^4 (-4+z)^2}
\eea
 We see here that for $\xi=0$  one can not meet $0<1-n_s \sim (12/ \sqrt[3]{9}) \ell^{4/3}<<1$ for $\ell >1$, whereas for $z \sim 4/33 (14 \pm \sqrt{97}$ (roots of the numerator of ($1-n_s$)) and having $\a$ quite large, one can satisfy ($0<1-n_s<<1, 0<r<<1$)). The red points in Fig (\ref{fig1}) represent acceptable points generated upon scanning the parameters as follows.
 \bea
\label{scanning-sceanarioI}
z \in [0.48,0.52], \ell \in [1,2], \varphi \in [1,20],\a \in [400,500].
\eea

%%%%%%%%%%%%%%%%%%%%%%%%%%%%%%%%%%%%%%%%%%
\section{Constant roll k-inflation. Case $\a <<1$}
In contrast to the preceding sections, we now take the perturbative limit $\a <<1$, and we work up to first order in $\a$. We shall consider a specific type of k-inflation called ``constant roll'' inflation, where one slow-roll parameter ($\eps_2$) related to the time second derivative of the inflaton is assumed constant equaling $\beta$. Following \cite{Odintsov_2019}, our model which has the following action:
\bea
\label{action-scenarioIII}
S &=& \int d^4x \sqrt{-g}/2 f(R,\chi,X)
\eea
where
\bea
\label{secanrioIII-details}
 X=\frac{1}{2}\partial_\mu \chi \partial^\mu \chi = \frac{\dot{\chi}^2}{2} &,& f(R,\chi,X)=\left( R-2X+4\a X^2 -2U\right)
\eea
will involve the slow-roll parameters defined as
\bea
\label{scenarioIII-eps-parameters}
\eps_1\equiv \frac{\dot{H}}{H^2}, \eps_2=\beta \equiv \frac{\ddot{\chi}}{H\dot{\chi}}&,&\eps_3 \equiv \frac{\dot{F}}{2HF}=0,\eps_4 \equiv \frac{\dot{E}}{2HE}
\eea
where
\bea
F=f_{,R}=1&,&E \equiv -\frac{F}{2X}(Xf_{,X}+2X^2f_{,XX})=1-12\a X
\eea
At the horizon crossing time instance, we have
\bea
\label{scenarioIII-eps-fields}
\eps_1 = -\frac{3}{4} \frac{\dot{\chi}^2+\a \dot{\chi}^4}{U(\chi)} &,& \eps_4 = \frac{6\sqrt{3}\a \dot{\chi} \ddot{\chi}}{\sqrt{U}(1+6\a \dot{\chi}^2)}
\eea
with real solutions given by
\bea
\label{scenarioIII-fields}
\dot{\chi}=\frac{6(\b+1)(\b+3)4\a U-(81\Delta+9\sqrt{S})^{2/3}}{3^{11/6}(\b+1)4\a \sqrt{U} (81\Delta+9\sqrt{S})^{1/3}} &,& \ddot{\chi} = \b \sqrt{\frac{U}{3}} \\
S=(\b+1)^3 (4\a)^3 U^2 \left[81(\b+1)4\a U_{,\chi}^2+ \frac{8}{3} (\b+3)^3U\right] &,&
\Delta= 16(\b+1)^2\a^2  U_{,\chi} U
\eea
 The spectral parameters are given as
 \bea
  n_s=1+2\frac{2\eps_1-\eps_2+\eps_3-\eps_4}{1+\eps_1}=1+2\frac{2\eps_1-\b-\eps_4}{1+\eps_1}
 \\ r=4\left[\frac{\Gamma(3/2)}{\Gamma(3/2+\eps_2)2^{\eps_2}}c_A^{3/2+\eps_2}\frac{\sqrt{3}\dot{\chi}\sqrt{1+6\a \dot{\chi}^2}}{\sqrt{U}}\right]^2 \\
 c_A^2=\frac{f_{,X}}{f_{,X}+2Xf_{,XX}}=\frac{-1+2\a \dot{\chi}^2}{-1+6\a \dot{\chi}^2}
 \label{scenarioIII-ns-r}
 \eea
The free input parameters here are ($\a, \xi, V_0$) and ($\ell, h$), which a priori determine $\chi$, and also $\b$ of order unity expressing the constant roll condition. However, note that we need to express $U_{,\chi}$ using Eqs. (\ref{Ueffective-h}, \ref{h-chi}).
\bea
\label{h-chi-again}
U_{,\chi}=\frac{dU}{dh}\frac{dh}{d\chi} &,& \frac{dh}{d\chi} = \pm \sqrt{1-\xi e^{2\ell h}+\frac{8\a V_0 e^{-2\ell h}}{1-\xi e^{2\ell h}}}
\eea
and even in the case of MCtG ($\xi=0$), where we get an analytic expression of $\chi$ in terms of $h$:
\bea \chi= \int \frac{dh}{\sqrt{1+8\a V_0 e^{-2 \ell h}}} = \frac{e^{-\ell h} \sqrt{8\a V_0 + e^{2 \ell h}} \log(e^{\ell h} + \sqrt{8\a V_0 + e^{2 \ell h}})}{\ell \sqrt{1+8\a V_0 e^{-2 \ell h}}}\eea
 however one can not invert it so $U(\chi)$ is not obtained in a closed form.

\section{Viability of the constant roll k-inflation: case $\a <<1$}
  We show now the existence of viable points which fit the data. For this, we need a search strategy to reduce the number of input parameters, since our objective is limited to a proof of existence with no claim to exhaustive covering of all acceptable points, otherwise scanning over the formulae of Eqs. (\ref{scenarioIII-eps-fields}-\ref{scenarioIII-ns-r}), which are far from simple analytical formulae, is not a trivial task.

  Let us take the case of MCtG ($\xi=0$) which with our limit case ($\a<<1$) leads to
  \bea \chi \simeq h, \bar{U}=V\simeq U , U_{,\chi} \sim \ell U
  \eea
For the sake of showing the existence of acceptable solutions, if we assume the constant slow roll parameter $\eps_2=\b$ is quite small, so that to imply dropping of $\ddot{\chi}$, then $\eps_4$ is negligible as well. In order to meet the requirements ($\a <<1, \ell >1$), we shall scan over the one-dimensional sub-parameter space parameterized as
\bea
\label{scenarioIII-scale}
\a= \Lambda^{-n}, \ell = \Lambda^m, V_0 = \Lambda, \chi\sim h = \ell^{-1} = \Lambda^{-m}
\eea
with ($n,m>0$). Noting that $e^{\ell h}$ is of order $\mathcal{O}(1)$ we get for $\Lambda$ large
\bea
\label{dot-chi-order}
\dot{\chi} &=& \frac{\mathcal{O}(\Lambda^{1-n})- \left( \mathcal{O}(\Lambda^{-2n+m+2}) + \mathcal{O}(\Lambda^{1-3n/2}) \sqrt{\mathcal{O}(\Lambda^{2+2m-n}) +\mathcal{O}(\Lambda)} \right)^{2/3}}{\mathcal{O}(\Lambda^{1/2-n}) \left( \mathcal{O}(\Lambda^{2+m-2n}) + \mathcal{O}(\Lambda^{1-3n/2}) \sqrt{\mathcal{O}(\Lambda^{2+2m-n}) +\mathcal{O}(\Lambda)}  \right)^{1/3}}
\eea
Then in order to get the following quantities small
\bea
\label{scale-ns-r}
1-n_s \approx -\frac{4\eps_1}{1+\eps_1} &,\eps_1 = -\frac{3}{4} \frac{\dot{\chi}^2 + \a \dot{\chi}^4}  {U},& r \approx  \frac{12 \dot{\chi}^2}{U}
\eea
we need to enforce
\bea
\label{condition-n-m}
0<n<1 &,& 0<m<\frac{1-n}{4}
\eea
%\mathcal{O}(\Lambda^{})
Numerically, we checked the viability of the model for vanishing and non-vanishing $\beta$ parameter. By taking the following six choices, the obtained points for the upper four (lower two) choices corresponding to vanishing (non-vanishing) constant-roll parameters, represented in Fig. (\ref{fig1}) by green (pink) dots, do fit the data:
\bea
\beta =0, \Lambda = 10^6, n=0.5, m=0.1 &\Rightarrow& (1-n_s,r)=(0.0280827,0.112981), \\
\beta =0, \Lambda = 10^{5.8}, n=0.5, m=0.1 &\Rightarrow& (1-n_s,r)=(0.032229,0.12977), \\
\beta =0, \Lambda = 10^{6.1}, n=0.5, m=0.1 &\Rightarrow& (1-n_s,r)=(0.0262134,0.10542), \\
\beta =0, \Lambda = 10^{4.44}, n=0.6, m=0.001 &\Rightarrow& (1-n_s,r)=(0.0298586,0.120145) \\
\beta =1, \Lambda = 10^{4.44}, n=0.6, m=0.1 &\Rightarrow& (1-n_s,r)=(0.0288759,0.0566994) \\
\beta =10, \Lambda = 3 \times 10^{4}, n=0.6, m=0.1 &\Rightarrow& (1-n_s,r)=(0.0288759,0)
\eea
which proves the viability of the model.
%%%%%%%%%%%%%%%%%%%%%%%%%%%
\section{Warm inflation variant}
As mentioned earlier, the varying coupling inflation variants generally call for new physics in order to treat the reheating process and to provide for an exit scenario. This problem can be addressed in warm inflation paradigm where the perturbations are generated thermally from a dissipative term characterized by a decay rate parameter $\Gamma$, which is sufficiently strong compared to Hubble parameter $H$ charcterized by the ratio:
\bea Q=\frac{\Gamma}{3H}\eea
 Here the radiation is close to thermal equilibrium, and both the particle production rate and dissipation rate are controlled by $\Gamma$. The radiation takes place in parallel to the slow roll regime, and no need for a specific exit scenario.

We readdress our Bekenstein-like scenario within the warm inflation paradigm assuming non-MCtG and $f(R)$ gravity embodied in the potential of Eq. \ref{Ueffective-h}, where upon putting $\a=0=\xi$ we switch back to the original scenario of \cite{chamoun_jcap_2016}. We shall restrict also our study to the weak dissipative regime $Q<<1$, remembering that $Q=0$ corresponds to the cold inflation.

The temperature during inflation is given by \cite{ref1,ref2}
\bea
\label{warm_T}
T &=& \left(\frac{\Gamma_0 U^2_\vphi}{36 H^3 C_\g}\right)^{1/3}
\eea
where
\bea
\label{warm_def}
\Gamma = \Gamma_0 T &,& C_\g = \frac{\pi^2}{30} g_*
\eea
with $\vphi$ the canonical inflaton field, and we shall always take $g_*=228.75$, representing the number of relativistic degrees of freedom of radiation of created massless modes, evaluated within minimal supersymmtric standard model at temperatures higher than the electroweak phase transition. In order to compute the derivatives with respect to $\vphi$ in terms of the derivatives with respect to $h$, we as usual use Eqs. (\ref{derivatives}).

Using the approximation
\bea H &=& \sqrt{\frac{U}{3}},
\eea
we have the slowwroll parameters given by
\bea
\label{warm_slowroll}
\eps_V = \frac{1}{2} \left( \frac{U_\vphi}{U}\right)^2 &,\beta_V = \left( \frac{\Gamma_\vphi U_\vphi}{\Gamma U}\right),& \eta_V = \left(\frac{U_{\vphi\vphi}}{U}\right),
\eea
Two parameters interfere to represent corrections due to the non-trivial occupation number ($n_*$) and to thermal effects ($\omega$) given by:
\bea
\label{warm_n-w}
n_* = \frac{1}{e^{\frac{H}{T}}-1} &,& \omega = \frac{2\pi \Gamma_0 T^2}{3H^2}
\eea
and, finally, we get, in the limit $\omega <<1$, the expressions for the observables:
\bea
n_s-1 = -6 \eps_V + 2 \eta_V + \omega \left(\frac{15 \eps_V -2 \eta_V -9 \beta_V}{4}\right)
&,& r= \frac{16 \eps_V}{(1+Q)^2(1+2n_* + \omega)}
\eea

Numerically, we find that upon switching off modification of gravity (i.e. $\a = 0 = \xi$), the value of $r$ is generically large, and one  needs to fine tune and adjust the parameters in order to find acceptable points, whereas switching on $\a$ helps generically to reduce $r$ and one can fit the data easier. In Fig (\ref{fig1}) we designate two points in yellow and other two ones in blue sky fitting the data corresponding to $\a =0$ and $\a \neq 0$ respectively,  with the following choice of parameters.
\bea
\a =0, \xi = 0, \Gamma_0 = 41.02 \times 10^{10^{-7}}, V_0 = 2, \ell = 1.5, h = 11 \Rightarrow (n_s,r)=(0.965929,0.105719), \\
\a =0, \xi = 0, \Gamma_0 = 41.04 \times 10^{10^{-7}}, V_0 = 2, \ell = 1.5, h = 11 \Rightarrow (n_s,r)=(0.973216,0.105702), \\
\a =1000, \xi = 0, \Gamma_0 = 1, V_0 = 1, \ell = 1, h = 1.05 \Rightarrow (n_s,r)=(0.964472,0.0202682), \\
\a =1000, \xi = -0.002, \Gamma_0 = 1, V_0 = 1, \ell = 1, h = 1 \Rightarrow (n_s,r)=(0.963546,0.0202916),
\eea

%%%%%%%%%%%%%%%%%%%%%%%%%%%%%%%%%%%%%%%%%%
\section{Summary  and Conclusion}
We continued in this letter the work of \cite{chamoun_ijmpa_2021} on the inflationary model generated by varying strong coupling constant, and studied here the effect of the quadratic kinetic energy term which appears upon introducing an $f(R)$ gravity, represented by an $\a R^2$-term in the pure gravity Lagrangian. We investigated in Palatini formalism two extreme cases corresponding first to $(\a >>1)$, which represents thus a highly non-canonical k-inflation, and second to $(\a <<1)$ where we kept terms to first order in $\a$ and examined a specific type of the k-inflation, namely the constant roll inflation. In both cases, we showed the viability of the model for some choices of the free parameters in regards to the spectral parameters ($n_s, r$) when compared to the results of Planck 2018 separately and combined with other experiments. However, the k-inflation required a non-MCtG and fine-tuned adjustment in order to accommodate data, whereas the constant-roll is able to accommodate data even in MCtG situation irrespective of the value of the constant roll parameter $\beta$. This amendment of inflationary models, which were thought before not to fit the data, by assuming $f(R)$ gravity and/or non-minimal coupling to gravity is a strong hint that this may be applicable to inflationary models other than the one studied in this work.

Finally, we readdressed the same model, $\grave{a}$ la Bekenstein within warm inflation scenario, which potentially is devoid of the exit scenario complications. In a specific limit, the weak limit corresponding to small parameters $Q$ and $\omega$, the model is able to accommodate data especially when supplemented with $f(R)$ gravity.

\vspace{6pt}
\section*{{\large \bf Acknowledgements}}
N. Chamoun acknowledges support from ICTP-Associate program, and from the Alexander von Humboldt Foundation, and is grateful for the hospitality of the Bethe Center for Theoretical Physics at Bonn University.

% The following MDPI journals use author-date citation: Arts, Econometrics, Economies, Genealogy, Humanities, IJFS, JRFM, Laws, Religions, Risks, Social Sciences. For those journals, please follow the formatting guidelines on http://www.mdpi.com/authors/references
% To cite two works by the same author: \citeauthor{ref-journal-1a} (\citeyear{ref-journal-1a}, \citeyear{ref-journal-1b}). This produces: Whittaker (1967, 1975)
% To cite two works by the same author with specific pages: \citeauthor{ref-journal-3a} (\citeyear{ref-journal-3a}, p. 328; \citeyear{ref-journal-3b}, p.475). This produces: Wong (1999, p. 328; 2000, p. 475)

%%%%%%%%%%%%%%%%%%%%%%%%%%%%%%%%%%%%%%%%%%
%% for journal Sci
%\reviewreports{\\
%Reviewer 1 comments and authors’ response\\
%Reviewer 2 comments and authors’ response\\
%Reviewer 3 comments and authors’ response
%}
%%%%%%%%%%%%%%%%%%%%%%%%%%%%%%%%%%%%%%%%%%

\end{document}